\documentclass[aps,prb,showpacs,twocolumn,citeautoscript,superscriptaddress]{revtex4}
\usepackage{amsmath}
\usepackage{graphicx}
\usepackage{dcolumn}
\usepackage{float}

\begin{document}
\title{Heavy fermion and Kondo lattice behavior in the itinerant ferromagnet CeCrGe$_{3}$}
\author{Debarchan Das}
\affiliation{Department of Physics, Indian Institute of Technology, Kanpur 208016, India}
\author{T. Gruner}
\affiliation{Max-Planck Institute for Chemical Physics of Solids, 01187 Dresden, Germany}
\author{H. Pfau}
\affiliation{Max-Planck Institute for Chemical Physics of Solids, 01187 Dresden, Germany}
\author{U. B. Paramanik}
\affiliation{Department of Physics, Indian Institute of Technology, Kanpur 208016, India}
\author{U. Burkhardt}
\affiliation {Max-Planck Institute for Chemical Physics of Solids, 01187 Dresden, Germany}
\author{C. Geibel}
\affiliation {Max-Planck Institute for Chemical Physics of Solids, 01187 Dresden, Germany}
\author{Z. Hossain}
\email{zakir@iitk.ac.in}
\affiliation{Department of Physics, Indian Institute of Technology, Kanpur 208016, India}
\affiliation {Max-Planck Institute for Chemical Physics of Solids, 01187 Dresden, Germany}

\date{\today}

\begin{abstract}

Physical properties of polycrystalline CeCrGe$_{3}$ and LaCrGe$_{3}$ have been investigated by x-ray absorption spectroscopy, magnetic susceptibility $\chi(T)$, isothermal magnetization M(H), electrical resistivity $\rho(T)$, specific heat C($T$) and thermoelectric power S($T$) measurements. These compounds are found to crystallize in the hexagonal perovskite structure (space group \textit{P6$_{3}$/mmc}), as previously reported.  The $\rho(T)$, $\chi(T)$ and C($T$) data confirm the bulk ferromagnetic ordering of itinerant Cr moments in LaCrGe$_{3}$ and CeCrGe$_{3}$ with $T_{C}$ = 90~K and 70~K respectively. In addition a weak anomaly is also observed near 3~K in the C($T$) data of CeCrGe$_{3}$. The T dependences of $\rho$ and finite values of Sommerfeld coefficient $\gamma$ obtained from the specific heat measurements confirm that both the compounds are of metallic character. Further, the $T$ dependence of $\rho$ of CeCrGe$_{3}$ reflects a Kondo lattice behavior. An enhanced $\gamma$ of 130~mJ/mol\,K$^{2}$ together with the Kondo lattice behavior inferred from the $\rho(T)$ establish CeCrGe$_{3}$ as a moderate heavy fermion compound with a quasi-particle mass renormalization factor of $\sim$ 45.

\end{abstract}


\maketitle

\section{INTRODUCTION}

Superconductivity in iron pnictide was discovered at the boundary of magnetically ordered and non-ordered state of Fe-magnetism. The spin-density-wave type antiferromagnetic order in RFeAsO (R =La, Ce, Pr, Nd, Sm) and AFe$_{2}$As$_{2}$ (A=Ca, Ba, Sr, Eu) could be suppressed by carrier doping\cite{Hiroki, Athena, Anupam} or by application of chemical/externally applied pressure\cite{Rotter, Park, Ren}. One may expect to find similar interesting results by tuning the magnetic state of Cr-moments as well. Recently Cr-magnetic ordering at relatively low temperature has been reported in RCrGe$_{3}$ system where R stands for rare-earth elements \cite{Bie}. On the other hand, Ce-based ternary germanide, RTGe$_{3}$ (R = rare-earth elements and T = $3d, 4d, 5d$ elements) system exhibits some interesting magnetic properties. For instance, CeRhGe$_{3}$ shows a weak Kondo effect along with three magnetic phase transitions \cite{Muro}. Ce moments order antiferromagnetically below $T_{N}$ = 5.5~K in CeNiGe$_{3}$ [9] which exhibits pressure induced superconductivity \cite{Nakashima} below 0.48~K in a wide pressure range from 4 to 10 GPa. Furthermore, a complex magnetic phase diagram is found in CeCoGe$_{3}$ [11] which also exhibit superconductivity under the application of external pressure \cite{Settai}.

The study of the anomalous physical properties close to the borderline between magnetically ordered and nonmagnetic ground states is currently of intense interest in the contemporary condensed matter physics research. This point of instability achieved by non-thermal tuning parameters between two stable phases of matter is called Quantum Critical Point (QCP). Interestingly, a quantum critical behavior is evidenced in doped CeCoGe$_{3}$ [13,14]. The situation may, however, get more complicated in CeCrGe$_{3}$ due to the presence of Cr-magnetic ordering. Very recently the suppression of Cr ferromagnetic ordering has been reported in LaV$_{x}$Cr$_{1-x}$Ge$_{3}$ system \cite{Lin} with a possibility for the presence of a QCP in this system. Such possibility also exists in the corresponding Ce-system and this motivates us to study CeCrGe$_{3}$ in detail.

The compounds RCrGe$_{3}$ are reported to crystallize in the hexagonal perovskite structure (space group \textit{P6$_{3}$/mmc}) \cite{Bie}. A preliminary investigation based on magnetization measurements of CeCrGe$_{3}$ reports ferromagnetic ordering of Cr-moments below 66~K; much less is known about the Ce-magnetism and its interaction with Cr-moments in this compound. We herein report comprehensive study of low temperature properties of CeCrGe$_{3}$ by means of x-ray absorption spectroscopy (XAS), magnetic susceptibility $\chi(T)$, isothermal magnetization M(H), electrical resistivity $\rho(T)$, specific heat C($T$) and thermoelectric power S($T$) measurements. In order to explore the role of $4f$-electron we have also measured the properties of the corresponding $4f^{0}$-electron analog LaCrGe$_{3}$ . Our results corroborate that Cr ions order ferromagnetically below 70~K in CeCrGe$_{3}$ and Ce-ions possess a stable 3+ valance state in this compound. Our observation of a large value of Sommerfeld coefficient $\gamma$, obtained from the specific heat measurement, even in the magnetically ordered state is quite remarkable in this context. The Kondo lattice type of the resistivity and the large $\gamma$  tempt us to believe that CeCrGe$_{3}$ is a moderate heavy fermion system.

\section{Experimental Details}

Polycrystalline samples of LaCrGe$_{3}$ and CeCrGe$_{3}$ were prepared by arc melting the high purity elements, La and Ce pieces (99.9\%, Alfa-Aesar), Cr and Ge chips (99.99\% Sigma Aldrich), taken in stoichiometric ratio on water cooled copper hearth under an argon atmosphere. The samples were flipped after each melting and were melted several times to ensure homogeneity.  In order to homogenize and remove the impurity phases the melted buttons were wrapped in Ta foil and annealed in an evacuated quartz tube at 900$^\circ$C for one week. The phase purities of the annealed samples were checked by powder x-ray diffraction (XRD) using Cu-$K_\alpha$ radiation and metallographic examinations. The absorption spectra of CeCrGe$_{3}$ at the Ce-$L_{III}$ edge (E = 5723~eV) have been recorded in transmission mode at the EXAFS beam line A1 of the Hamburg synchrotron radiation facility (HASY-LAB at DESY) using the Si $(111)$ double crystal monochromator. In order to perform the XAS study finely grounded/powdered sample of $\sim$ 12 mg with particle size smaller than $20  \mu$m was mixed with B$_4$C and PE powder and pelletized to a 10~mm disc. The absorption behavior has been determined in the energy range from 5500~eV up to the Cr K absorption edge (5989~eV) with a minimal step size of $\Delta$E =0.25~eV near the Ce-$L_{III}$ edge (5723~eV). The pellet was mounted in an Oxford gas flow cryostat to realize temperatures down to the boiling point of liquid helium. The spectra showed an absorption edge step of $\Delta \mu = 0.8$ which represents the difference of the low and high energy background functions at the absorption edge. The (Ce$^{3+}$)-reference CePO$_{4}$ was measured simultaneously at ambient temperature and was used for energy calibration. Evaluation of spectra, energy calibration and normalization have been performed with the Athena module of the Horae software package \cite{Athena1}. Magnetic measurements were carried out using a commercial Superconducting Quantum Interference Device (SQUID) magnetometer (MPMS Quantum Design). Electrical resistivity measurements were carried out in the temperature range 2-300 K using ac transport option of physical property measurement system (PPMS, Quantum Design). The specific heat was measured by relaxation method in PPMS.  The thermoelectric power (TEP) was measured using thermal transport option of PPMS.

\section{RESULTS AND DISCUSSION}

The crushed polycrystalline samples were characterized by x-ray diffraction with Cu-$K_\alpha$ radiation to determine the single phase nature and the crystal structure. Powder x-ray data were analyzed by Rietveld refinement using FullProf software\cite{Rodriguez}. The refinement reveals the single phase nature of the samples. The lattice parameters for CeCrGe$_{3}$ and LaCrGe$_{3}$, given in Table I, are in good agreement with those reported in the literature\cite{Bie}. The room temperature XRD pattern and Rietveld refinement profile are shown in Fig 1. Compositional homogeneity was checked by performing metallographic examinations like Scanning electron microscopy (SEM) and Energy dispersive x-ray spectroscopy.
\begin{figure}
\includegraphics[width=8.5cm, keepaspectratio]{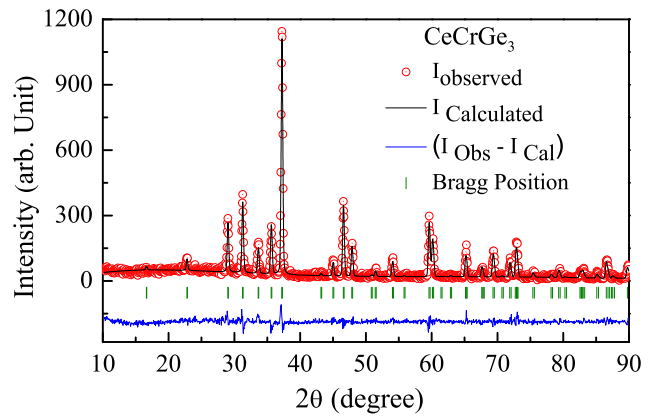}
\caption{\label{fig:xray} (Color online) The powder x-ray diffraction pattern of CeCrGe$_{3}$ recorded at room temperature. The solid line through the experimental points is the Rietveld refinement profile calculated for the BaNiO$_{3}$ -type hexagonal perovskite structure (space group \textit{P6$_{3}$/mmc}). The short vertical bars mark the Bragg peak positions. The lowermost curve represents the difference between the experimental and calculated intensities.}
\end{figure}
\begin{table}
\caption{\label{tab:XRD} Table I: Crystallographic data for RCrGe$_{3}$ (R = La,Ce) obtained from the structural Rietveld refinement of powder XRD data.}
\begin{ruledtabular}
\begin{tabular}{c c c c}
Structure &\multicolumn{3}{l} {BaNiO$_{3}$-type Hexagonal} \\
Space group & \textit{P6$_{3}$/mmc} \\ \\
Lattice parameters & LaCrGe$_{3}$ & CeCrGe$_{3}$ \\

 $a$ (\AA)  & 6.198  & 6.150  \\
 $c$ (\AA)  & 5.765  & 5.719  \\
 $V_{cell}$ (\AA$^3$)& 191.79 & 187.32  \\

\end{tabular}
\end{ruledtabular}
\end{table}
\begin{figure}
\includegraphics[width=8.5cm, keepaspectratio]{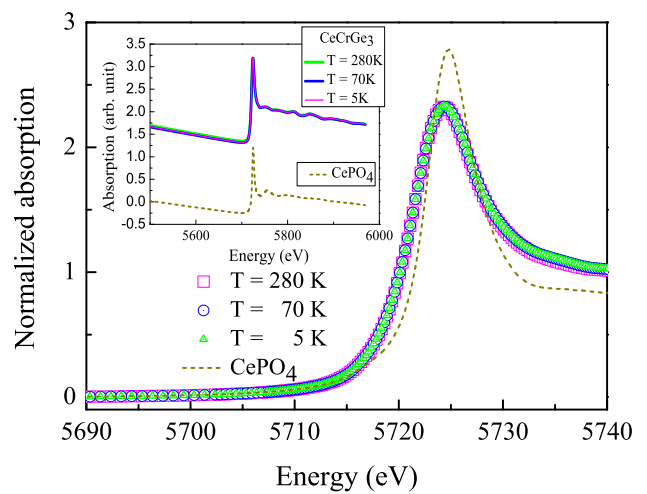}
\caption{\label{fig:XAS}(Color online) Ce-$L_{III}$ edge x-ray absorption of CeCrGe$_{3}$ at 280~K, 70~K and 5~K in comparison to the (Ce3+) reference spectrum of CePO$_{4}$. (Inset: un-normalised spectra).}
\end{figure}

To elucidate the valence state of Ce in CeCrGe$_{3}$ we performed Ce-$L_{III}$ edge x-ray absorption study. The spectra of CeCrGe$_{3}$ and the (Ce$^{3+}$)- reference compound CePO$_{4}$  show one single white line at the Ce-$L_{III}$  edge with a typical broadening in the spectra of an intermetallic compound. Neither the spectra of CeCrGe$_{3}$ nor its derivative indicate an additional (Ce$^{4+}$) contribution which would lead to an additional absorption line at approximately 5730~eV. The spectrum is temperature independent down to T = 5~K and consequently the (Ce$^{3+}$) valence state remains unchanged in the examined range 5~K $<$ T $<$ 280~K.

\begin{figure}[htb!]
\includegraphics[width=8.5cm, keepaspectratio]{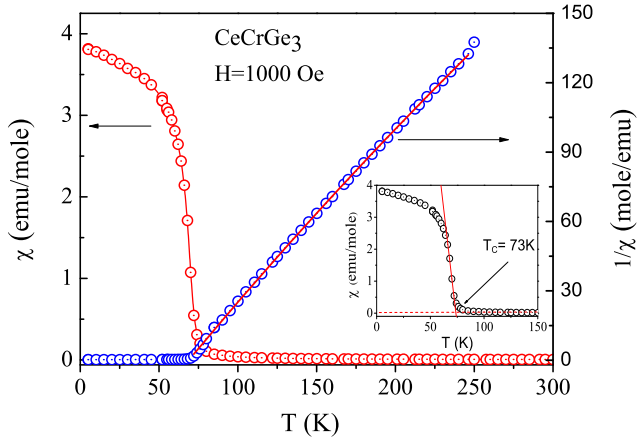}
\caption{\label{fig:MT}(Color online) Field-cooled (FC) dc magnetic susceptibility and inverse susceptibility of CeCrGe$_{3}$. Inset: Magnetic susceptibility in the temperature range 2 K - 150 K. The Magnetic ordering temperature is represented by the meeting point of the solid and dotted lines. }
\end{figure}

The temperature dependence of the magnetic susceptibility $\chi(T)$ of CeCrGe$_{3}$ is shown in Fig. 3 in the temperature range 2-300~K under an applied field of 0.1~T. At high temperature $\chi(T)$ follows the modified Curie-Weiss behavior, $\chi(T)= \chi_{0}+ C/(T - \theta_{P})$ where $\chi_{0}$ is temperature independent susceptibility, C is the Curie constant and $\theta$$_{P}$ is the Weiss temperature. Fig. 3 also displays the inverse magnetic susceptibility as a function of temperature along with the fitting of $\chi^{-1}$(T) data with the modified Curie-Weiss behavior in the high temperature range.  The effective moment $\mu_{eff}$ values calculated from the Curie constants are found to be 2.66~$\mu_{B}$ and 3.18~$\mu_{B}$ for LaCrGe$_{3}$ and CeCrGe$_{3}$ respectively. The observation of this increased effective moment value for CeCrGe$_{3}$ is consistent with the trivalent state of Ce-ions revealed by XAS measurement. It is to be noted in this context that these values are higher than those reported in Ref. 7. For LaCrGe3 a value of 2.5~$\mu_{B}$ has been reported recently by Lin et al. [15] which is nearly same as the value obtained by us. For CeCrGe$_{3}$ $\mu_{eff}$ is higher than the calculated value for free Ce$^{3+}$: 2.54~$\mu_{B}$. As it is confirmed by XAS study that Ce in CeCrGe$_{3}$ possesses a stable 3+ valance state with a theoretical value of $\mu_{eff}$ = 2.54~$\mu_{B}$, Ce contribution to the Curie constant (${C}$) is $C_{Ce}$= 0.81 emu K/mol. So, now by subtracting $C_{Ce}$ from ${C}$ (=1.26 emu K/mol) we can determine the Cr contribution to the Curie constant, $C_{Cr}$, which gives an effective moment value of $\mu_{eff}$(Cr)= 1.81~$\mu_{B}$[see Ref. No 18 for analysis]. This $\mu_{eff}$(Cr) value is less than that for Cr$^{4+}$ (2.8~$\mu_{B}$) or Cr$^{3+}$ (3.8~$\mu_{B}$). The reduced Cr moment suggests an itinerant character of Cr magnetism. The positive value of Weiss temperature $\theta$$_{P}$ suggests the presence of dominant ferromagnetic interactions in the system. The Curie temperatures $T_C$ were derived by extrapolating the maximum slope (d$\chi$/d$T$) to zero as shown in the inset of Fig. 3. The effective moments $\mu_{eff}$, Weiss temperatures $\theta$$_{P}$ and Curie temperatures $T_C$ are summarized in Table II.

Fig. 4 represents the magnetic isotherms M(H) of CeCrGe$_{3}$ measured at 1.8~K and 300~K under applied magnetic fields up to 7~T. In the ordered state M(H)at 1.8~K shows a rapid increase up to a value of $\sim$ 0.6~$\mu_{B}$ at low field followed by slow increase beyond 0.2~T. This may be due to possible canted ferromagnetic ordering of Cr-moments as revealed in LaCrGe$_{3}$ by neutron diffraction study \cite{Cadogan}. At 1.8~K magnetization reaches a value of 0.8~$\mu_{B}$/f.u. at 7~T for CeCrGe$_{3}$ which is lower than the value 1.3~$\mu_{B}$/f.u. at 7~T obtained for LaCrGe$_{3}$. The reduced value for CeCrGe$_{3}$ could arise due to crystal field effect/Kondo effect/ interaction between the Ce and Cr sublattice. At 300~K M(H) is almost linear in H as expected in the paramagnetic state.
\begin{figure}[htb!]
\includegraphics[width=8.5cm, keepaspectratio]{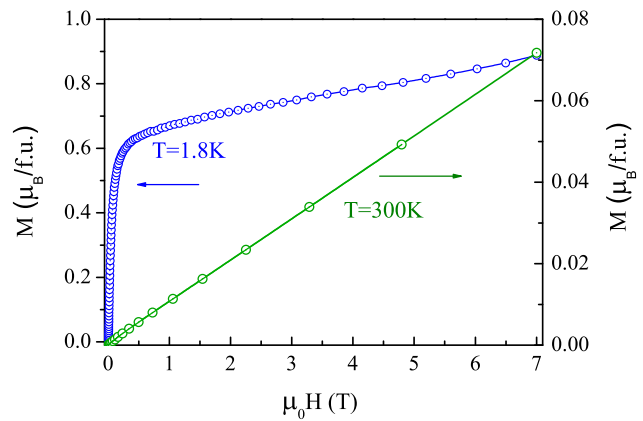}
\caption{\label{fig:Isothermal magnetization}(Color online) Magnetic field dependence of isothermal magnetization data, M(H) of CeCrGe$_{3}$ measured at constant temperatures 1.8~K and 300~K.}
\end{figure}

The electrical resistivity $\rho(T)$ of LaCrGe$_{3}$ and CeCrGe$_{3}$ measured in the temperature range 2-300~K is shown in Fig. 5(a).  While LaCrGe$_{3}$ exhibits a typical metallic behavior with residual resistivity $\rho_0$ $\sim $ 65~$\mu~\Omega$~cm at 2~K and residual resistivity ratio RRR = $\rho_{300\,{K}}/\rho_{2\,{K}} \sim  4.3$, the electrical resistivity profile of CeCrGe$_{3}$ interestingly exhibits Kondo like behavior. For CeCrGe$_{3}$, the resistivity increases with decreasing temperature down to 70~K followed by a drastic fall of electrical resistivity below 70~K as a consequence of reduction of spin disorder resistivity due to Cr-moment ordering. Below 30~ K the resistivity again increases similar to the case of Kondo insulator/semiconductor. Here, the temperature at which the resistivity anomaly occurs has been denoted by $T_{c}^{(\rho-T)}$ and is summarized in Table II. The magnetic contribution to the resistivity $\rho_{mag}$ obtained by subtracting the resistivity of LaCrGe$_{3}$ from that of CeCrGe$_{3}$ is depicted in Fig. 5(b). The temperature variation of $\rho_{mag}$ follows -ln$T$ behavior in two different temperature regions which we believe is associated with the presence of Kondo effect as seen in many Ce-based Kondo Lattice/Heavy fermion systems. These two regions are separated by the FM transition around 70~K. A sharp increase of $\rho_{mag}$ (Ce-La) below 86~K is an artificial effect of the subtraction process because $T_{C}$ = 88~K of the La-system is larger than $T_{C}$ = 68~K of the Ce-system.
\begin{figure}[htb!]
\includegraphics[width=8.5cm, keepaspectratio]{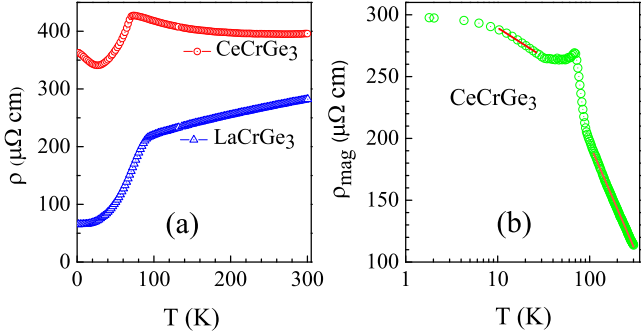}
\caption{\label{fig:Electrical resistivity}(Color online) (a) Electrical resistivity of LaCrGe$_{3}$ and CeCrGe$_{3}$, (b) Magnetic contribution to the electrical resistivity. Solid lines are fit to the equation $\rho_{mag}$= $\rho_{0}$ - C$_{k}$ ln$T$}
\end{figure}
We have analyzed the data of $\rho_{mag}$ using the expression:
\begin{equation}
\rho_{mag} = \rho_{0} - C_{k} ln T
\label{eq:C}
\end{equation}
\noindent where $\rho_{0}$ is the spin disorder resistivity and C$_{k}$ is the Kondo coefficient. The fitting to this equation at two different temperature regions are displayed by the solid lines in Fig 5(b). The low temperature (10~K~-~25~K) -ln$T$ behaviour represents the Kondo scattering in the crystal field ground state whereas -ln$T$ dependence in the high temperature (110~K~-~300~K) part is due to Kondo effect in the excited multiplets \cite{Cornut}. The high temperature ln$T$ dependence is one of the characteristic features of the dense Kondo systems.

\begin{figure}
\includegraphics[width=8.5cm, keepaspectratio]{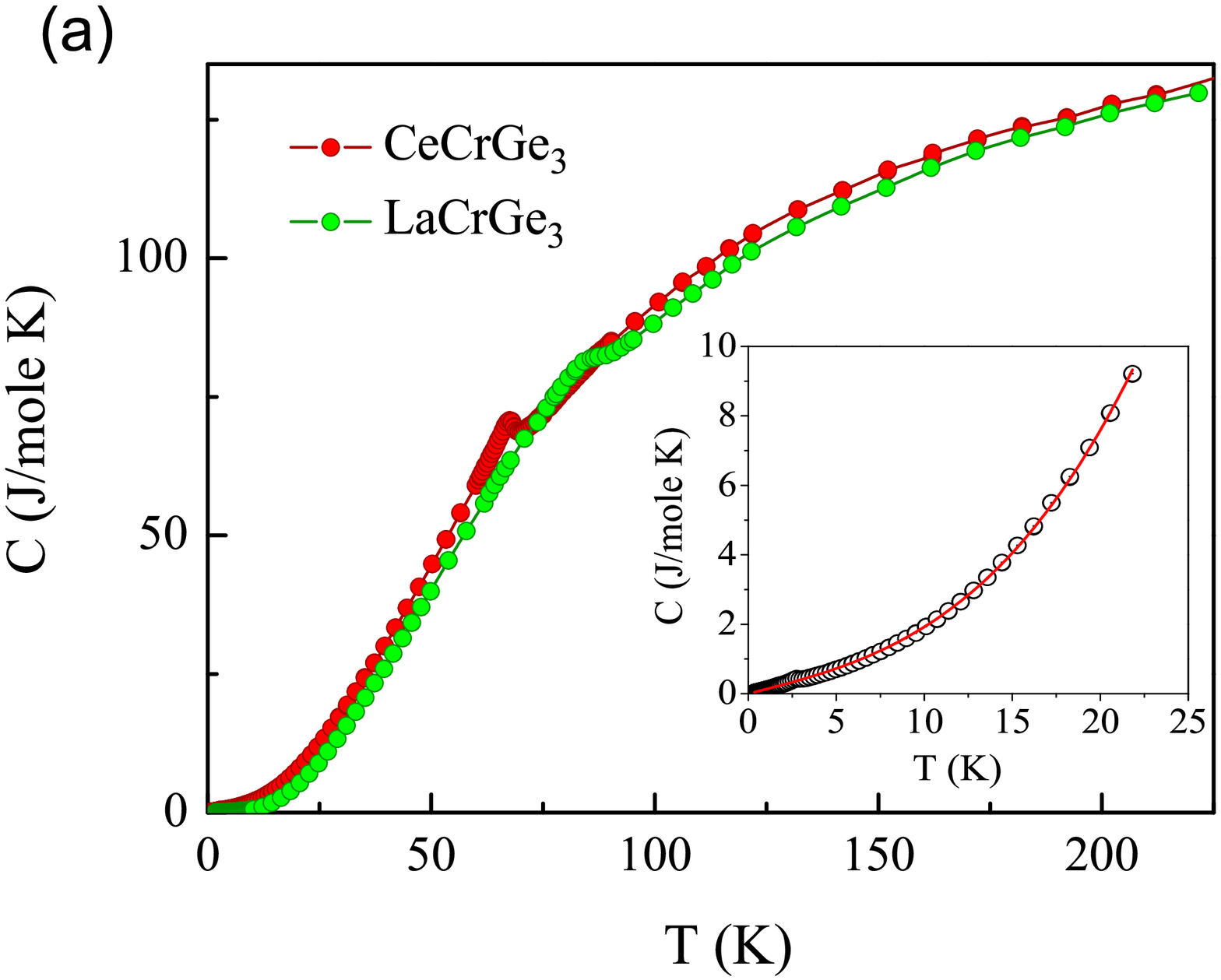}
\includegraphics[width=8.5cm, keepaspectratio]{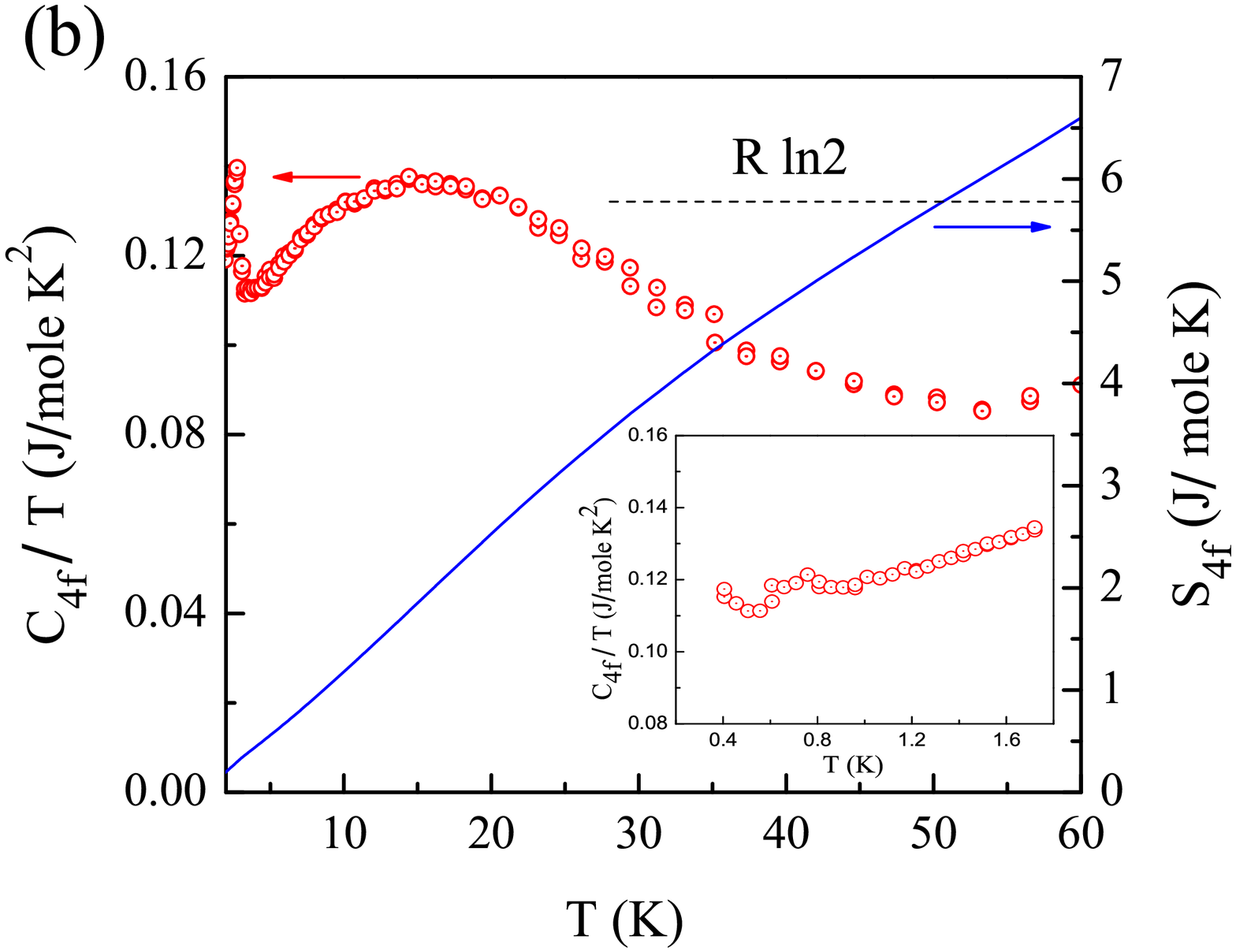}
\caption{\label{fig:Heat capacity}(Color online) (a) Specific heat of RCrGe$_{3}$ as a function of $T$. Inset shows the low temperature fit to the equation $C(T) = \gamma T + \beta T^{3}$ for CeCrGe$_{3}$. (b)  C$_{4f}$ and magnetic entropy (S$_{4f}$) as a function of temperature. Inset: C$_{4f}$ in the temperature range 0.4~K to 1.8~K. }
\end{figure}

Sometimes Kondo system can have large value of Sommerfeld coefficient $\gamma$. In order to get the information about the linear term $\gamma$ in CeCrGe$_{3}$ system exhibiting Kondo like behavior, we have performed specific heat measurements. Fig. 6(a) represents the temperature dependence of specific heat C($T$) of RCrGe$_{3}$, for R = La and Ce. The anomalies at $\sim$~84~K and 67~K as seen in C($T$) data for LaCrGe$_{3}$ and CeCrGe$_{3}$ respectively are consistent with the anomalies observed in temperature dependence of magnetic susceptibility as well as in electrical resistivity. The bulk nature of ferromagnetic ordering as evidenced by magnetic measurements and reported in the literature \cite{Bie,Lin} is confirmed by the anomaly in C($T$) data at 84~K and 67~K for LaCrGe$_{3}$ and CeCrGe$_{3}$ respectively. The low temperature specific heat data below 20~K can be fitted with the equation,

\begin{equation}
C(T) = \gamma T + \beta T^{3}
\label{eq:C}
\end{equation}
\noindent where $\gamma T$ is the electronic contribution to the specific heat, $\beta T^{3}$ is the phononic contribution to the specific heat [inset of the Fig. 6(a)]. The value of Debye temperature $\Theta_{D}$ can be estimated from $\beta$ using the relation \cite{Kittel},
\begin{equation}
\Theta_{D} = \left( \frac{12 \pi^{4} n R}{5 \beta} \right)^{1/3}
 \label{eq:Debye-Temp}
\end{equation}
\begin{table*}[ht!]
\centering
\caption{\label{tab:Table2} Summary of magnetization, electrical resistivity and specific heat data.}
%\begin{center}
\setlength{\tabcolsep}{6pt}
\begin{tabular}{c c c c c c c}
\hline
\hline
Sample &$\mu_{eff}(\mu_{B}$)  &  $\theta_{P}$(K)  &  $T_{c}^{(mag)}$(K) & $T_{c}^{(\rho-T)}$(K) &  $\Theta_{D}$(K) & $\gamma$(mJ/mol\,K$^{2}$)\\
\hline

LaCrGe$_{3}$     &2.66  &  87  &  88  &  88  &  253 & 2.89\\ [1ex]
CeCrGe$_{3}$   &3.18  &  67  &  73  &  68  &  250  & 130\\ [0.5ex]
\hline
\hline
\end{tabular}
%\end{center}
\end{table*}

\noindent where R is the molar gas constant and n = 5 is the number of atoms per formula unit (f.u.). We have listed the obtained values of $\gamma$ and $\Theta_{D}$ for the two compounds in Table II. The Sommerfeld coefficient ($\gamma$) for CeCrGe$_{3}$ is found out to be 130 ~mJ/mol\,K$^{2}$. This value is very large compared to the $\gamma$ value 2.9 ~mJ/mol\,K$^{2}$ obtained for LaCrGe$_{3}$ having 4$f^0$ configuration. This $\gamma$ value of CeCrGe$_{3}$ is comparable with that found in case of moderate heavy fermion system like CeFeGe$_{3}$ [21], Ce$_{2}$Rh$_{3}$Ge$_{5}$ [22], CeCoGe$_{3}$ [11], YbNi$_{2}$B$_{2}$C [23] etc. Assuming that the quasi-particle mass renormalization in LaCrGe$_{3}$ is negligible, the renormalization factor for quasi-particle density of states due to 4f correlations in CeCrGe$_{3}$ can be obtained by the ratio $\gamma$(CeCrGe$_{3}$)/ $\gamma $(LaCrGe$_{3}$) which comes out to be $\sim$ 45. Hence, the quasi-particle mass in CeCrGe$_{3}$ can be roughly estimated as m*$\approx$ 45m$_{e}$, where m$_{e}$ represents the free electron mass. Although a band structure calculation or de Haas-van Alphen studies on single crystal of CeCrGe$_{3}$ is required for a better estimate of m*, the effective mass estimated in this way was found to be very close to the experimentally observed value for RRhSi$_{3}$ system (R= La and Ce) where LaRhSi$_{3}$ has $\gamma$ = 6~mJ/mol\,K$^{2}$ [25] and CeRhSi$_{3}$ has $\gamma$ = 120~mJ/mol\,K$^{2}$[26]. Thus the ratio of $\gamma$ gives m* = 20m$_{e}$ which is very close to the experimentally observed m* = 19~m$_{e}$ to 24~m$_{e}$ in de Haas-van Alphen studies on CeRhSi$_{3}$[27,28]. Thus, the observation of a typical Kondo lattice behavior in electrical resistivity and an enhanced $\gamma$ value suggest that CeCrGe$_{3}$ can be classified as a member of the heavy fermion system. The observation of heavy fermion behavior even in the presence of ferromagnetic order having large ordering temperature is remarkable in this system.

In order to obtain the Ce-$4f$ contribution to the specific heat (C$_{4f}$), we have subtracted the heat capacity of LaCrGe$_{3}$ from that of CeCrGe$_{3}$ which is displayed in a plot of C$_{4f}/T$ versus $T$, Fig 6(b). The first observation is that the $T$ dependence of the C$_{4f}/T$ value is very flat. Besides the small sharp anomaly at $T$ = 2.8~K there is only a weak and broad maximum at 15~K followed by a smooth decrease to higher $T$, which however amount to less than 30\% between 15~K and 50~K. The overall C$_{4f}/T$ value is enhanced, of the order of 120 ~mJ/mol\,K$^{2}$. For a stable trivalent Ce$^{3+}$ without Kondo interaction, one would expect a polarization of the Ce-moment in the exchange field produced by FM ordering of Cr, possibly followed by ordering of Ce on its own at lower temperatures. This would result in a clear Schottky anomaly due to Zeeman splitting, possibly followed by a sharp anomaly, and ending in a pronounced decrease to lowest T since the Ce system is then in a fully polarized state \cite{Jesche1, Jesche2}. This scenario does not fit at all to the experimental observations made here, namely, C$_{4f}/T$ remains large up to the lowest $T$ investigated, 0.4~K (inset Fig. 6(b)), much more similar to the expected behavior of the heavy fermion system. This large C$_{4f}/T$ value at lowest $T$ implies strong fluctuations extending down to $T$ = 0~K. Assuming these fluctuations are due to Kondo effect, C$_{4f}/T$ allows for an estimate of the Kondo temperature $T_{K}$ using the theoretical expression derived by Rajan et al \cite{Rajan, Satoh, Rajan2}
\begin{equation}
\left(\frac{C_{4f}}{T}\right)_{T=0}= \frac{(\nu - 1)\pi R} {6 T_0}
\label{eq:C}
\end{equation}
\noindent where R is the gas constant, $\nu$ is the degeneracy of the lowest level concerned and $T_0$ is a scaling temperature which is related to the Kondo temperature $T_K$ by $T_K$/$T_0$ = 1.290. Taking ground state degeneracy $\nu$ = 2, and  (C$_{4f}/T$)$_{T = 0}$ = 0.117~J/mol\,K$^{2}$, $T_K$ = 48~K and taking $\nu$ = 4 we get $T_K$ = 143~K.

Furthermore, the $4f$-entropy obtained from the integration of C$_{4f}/T$ increases continuously at least up to 50~K where it has reached a value close to Rln2 [Fig 6(b)]. Even above 50~K the magnetic entropy S$_{4f}(T)$ continues to increase with T, this implies that the energy splitting to first excited CEF level should be smaller than 150~K. However, the absence of saturation behavior towards Rln2 implies that this splitting is not well defined (resolved). The high $T_K$ obtained from the Sommerfeld coefficient provides a first simple explanation, since a strong correlation that is responsible for high gamma value would lead to a broadening of the CEF level of the same order of magnitude as the splitting. However, the exchange field due to the FM ordering of Cr moments provides a further mechanism, but the exchange field needed is huge, since it has to lead to Zeeman splitting of the order of 150~K. Inelastic neutron scattering experiments are required to have a clear idea about the CEF level scheme in this system.

\begin{figure}[htb!]
\includegraphics[width=9cm, keepaspectratio]{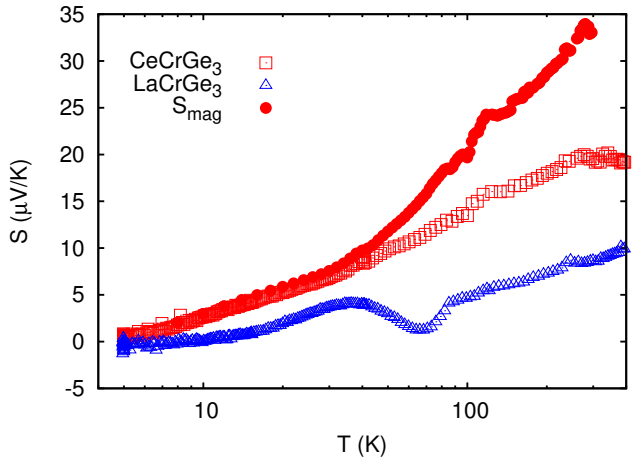}
\caption{\label{fig:thermopower}(Color online) Thermoelectric power of LaCrGe$_{3}$ and CeCrGe$_{3}$ as a function of temperature.}
\end{figure}

As a complementary probe for characteristics of the Kondo effect, we performed thermopower measurements. Fig.7 shows the thermopower S($T$) of LaCrGe$_{3}$ and CeCrGe$_{3}$ between 5 and 300~K. Additionally we calculated the magnetic contribution S$_{mag}(T)$ to the thermopower of CeCrGe$_{3}$ by the Gorter-Nordheim relation,
\begin{equation}
S\ast\rho = S_{mag}\ast\rho_{mag} + S_0\ast\rho_0
\label{eq:C}
\end{equation}
\noindent where S$_0$ and $\rho_0$ are the data for the isostructural La compound. Typically, Ce-based Kondo systems exhibit (a) enhanced thermopower values, which are 10-100 times larger than that of simple metals like Cu \cite{Zlatic}, and (b) a positive sign of S($T$) \cite{Miyake}. For well-defined CEF resonances and a low $T_K$ one also finds (c) two maxima: one around the Kondo temperature $T_K$ and the second around a temperature corresponding to roughly 0.3 to 0.6 of the CEF splitting \cite{Zlatic}. They reflect Kondo scattering on the ground state and on the excited CEF states, respectively. Our data show (a) indeed a very high value of both S$_{mag}(T)$ (35~$\mu$ V/K) and S($T$)(20~$\mu$ V/K) of CeCrGe$_{3}$ . Additionally we find (b) that S($T$) is positive in the whole temperature range. Both properties indicate that CeCrGe$_{3}$ is a Kondo system. Concerning (c) we, however, do not see any extrema in S$_{mag}(T)$, but a monotonic increase up to room temperature. Since we expect a relatively high Kondo temperature from our specific heat measurements and additionally do not find clear signatures of the CEF levels in the specific heat either, we are far from the conditions mentioned for (c). Indeed, it is known from theoretical calculations describing experiments e.g. on CeRu$_{2}$Ge$_{2}$, that the temperature dependence of the thermopower can vary drastically and can even become monotonic depending on the Kondo scale and the CEF splitting \cite{Zlatic}. In the LaCrGe$_{3}$ system we additionally find a broad hump between 30 and 60~K, which is believed to be related to the ordering of the Cr moments.

\section{\label{Conclusions} CONCLUSIONS}

We have provided experimental evidence for the presence of Kondo effect/heavy fermion behavior of the trivalent Ce compound CeCrGe$_{3}$ in presence of the ferromagnetic order due to itinerant Cr-moments. While the conclusive evidence of the trivalent nature was obtained from XAS, the presence of Kondo behavior was revealed by electrical resistivity, thermoelectric power and heat capacity data. The heavy fermion nature was revealed using the heat capacity measurement. Specific heat data show an enhanced $\gamma$~(130~mJ/mol\,K$^{2}$) value for CeCrGe$_{3}$ compared to 2.9~mJ/mol\,K$^{2}$, obtained for LaCrGe$_{3}$. Although magnetic order of Cr was known from earlier magnetization data, our heat capacity data confirms the bulk nature of the magnetic order. Hence, the compound CeCrGe$_{3}$, having a high $T_K$ ($\sim$100~K) and a trivalent integral Ce valence state at room temperature as well as at low temperature, appears to be a rare example of Ce based system exhibiting heavy fermion behavior even in the magnetically ordered state.

\section*{ACKNOWLEDGEMENTS}

We would like to thank Dr. V. K. Anand for useful discussion. This work is partially supported by Council of Scientific and Industrial Research (CSIR), New Delhi (Grant No. 80(0080)/12/ EMR-II).


\begin{thebibliography}{44}

\bibitem{Hiroki}
 Hiroki Takahashi, Kazumi Igawa, KazunobuArii, Yoichi Kamihara, Masahiro Hirano and Hideo Hosono; Nature, {\bf 453}, 376 (2008).

\bibitem{Athena}
 Athena S. Sefat, Rongying Jin, Michael A. McGuire, Brian C. Sales, David J. Singh, and David Mandrus; Phys. Rev. Lett. {\bf 101}, 117004 (2008).

\bibitem{Anupam}
 Anupam, P L Paulose, H S Jeevan, C Geibeland Z Hossain; J. Phys.: Condens. Matter {\bf 21} 265701 (2009).

\bibitem{Rotter}
Marianne Rotter, Marcus Tegel and  Dirk Johrendt; Nature Materials {\bf 101}, 107006 (2008).

\bibitem{Park}
Tuson Park, Eunsung Park, Hanoh Lee, T Klimczuk, E D Bauer, F Ronning and J D Thompson; J. Phys.: Condens. Matter {\bf 20} 322204, (2008).

\bibitem{Ren}
Zhi Ren, Qian Tao, Shuai Jiang, ChunmuFeng, Cao Wang, Jianhui Dai, Guanghan Cao, and Zhu'anXu. Phys. Rev. Lett. {\bf 102}, 137002 (2009).

\bibitem{Bie}
Haiying Bie,Oksana Ya. Zelinska, Andriy V. Tkachuk and Arthur Mar ; Chem. Mater {\bf 19} 4613 (2007).

\bibitem{Muro}
 Yuji Muro, DuhwaEom, Naoya Takeda and Masayasu Ishikawa. J. Phys. Soc. Jpn. {\bf 67} 3601 (1998).

 \bibitem{Manfrinetti}
 P. Manfrinetti, S.K. Dhar, R. Kulkarni, A.V. Morozkin, Solid State Communications  {\bf 135}, 444 (2005).

 \bibitem{Nakashima}
 M Nakashima, K Tabata, A Thamizhavel, T C Kobayashi, M Hedo, Y Uwatoko, K Shimizu, R Settai and Y\={O}nuki, J.Phys.: Condens. Matter {\bf 16} L255 (2004).

 \bibitem{Pecharsky}
 V. K. Pecharsky, O.-B. Hyun, and K. A. Gschneidner, Jr.;  Phys. Rev. B. {\bf 47} 11839 (1993).

 \bibitem{Settai}
 R. Settai, I. Sugitani, Y. Okuda, A. Thamizhavel, M. Nakashima, Y. \={O}nuki, H. Harima; Journal of Magnetism and Magnetic Materials {\bf 310} 844 (2007).

 \bibitem{Eom}
 Duhwa Eom, Masayasu Ishikawa, Jiro Kitagawa and Naoya Takeda; J. Phys. Soc. Jpn. Vol. {\bf 67} 2495 (1998).

 \bibitem{Alzamora}
 M. Alzamora, M. B. Fontes, J. Larrea J, H. A. Borges, E. M. Baggio-Saitovitch and S. N. Medeiros; Phys. Rev. B {\bf 76}, 125106 (2007).

 \bibitem{Lin}
 Xiao Lin, ValentinTaufour, Sergey L. Bud'ko, Paul C. Canfield; Phys. Rev. B {\bf 88}, 094405 (2013).

 \bibitem{Athena1}
 Athena, Artemis, Hephaestus: data analysis for X-ray absorption spectroscopy using IFEFFIT, J. Synchrotron Rad.12, (2005) 537.

\bibitem{Rodriguez}
 J. Rodr\'{i}guez-Carvajal, Physica B {\bf 192}, 55 (1993); Program Fullprof, LLB-JRC, Laboratoire L\'{e}on Brillouin, CEA-Saclay, France, (1996).

\bibitem{Jackson}
 D. D. Jackson, S. K. McCall, A. B. Karki and D. P. Young; Phys. Rev. B, {\bf 76}, 064408 (2007).


\bibitem{Cadogan}
 J. M. Cadogan, P. Lemoine, B. R. Slater, A. Mar and M. Avdeev, Solid State Phenom. {\bf 194}, 71 (2013).

\bibitem{Cornut}
 B. Cornut and B. Coqblin, Phys. Rev. B, {\bf 5}, 4541 (1972).

 \bibitem{Kittel}
C. Kittel, \emph{Introduction to Solid State Physics}, 8th edition (Wiley, New York, 2005).

\bibitem{Yamamoto}
Hiroshi Yamamoto, Masayasu Ishikawa, Katsuhiro Hasegawa and Junji Sakurai, Phys. Rev.B {\bf 52} 10136 (1995).

\bibitem{Hossain}
Z. Hossain, H. Ohmoto, K. Umeo, F. Iga, T. Suzuki, T. Takabatake, N. Takamoto and K. Kindo; Phys. Rev. B, {\bf 60},14 (1999).

\bibitem{Dhar}
S. K.  Dhar,  R. Nagarajan,  Z. Hossain,  E. Tominez,  C. Godart,  L.C. Gupta and R. Vijayaraghavan;  Solid State Communications,  {\bf 98},  11 (1996).

\bibitem{Anand}
V. K. Anand, A. D. Hillier, D. T. Adroja, A. M. Strydom, H. Michor, K. A. McEwen and B. D. Rainford; Phys. Rev. B, {\bf 83}, 064522 (2011)

\bibitem{Muro}
Yuji Muro, Duhwa Eom, Naoya Takeda and Masayasu Ishikawa ; J. Phys. Soc. Jpn. {\bf 67}, 3601(1998).

\bibitem{Terashima}
T. Terashima, M. Kimata, S. Uji, T. Sugawara, N. Kimura, H. Aoki and H. Harima; Phys. Rev. B, {\bf 78}, 205107 (2008).

\bibitem{Kimura}
N. Kimura, Y. Umeda, T. Asai, T. Terashima and H. Aoki; Physica B {\bf 294-295},280 (2001)

\bibitem{Rajan}
V. T. Rajan; Phys. Rev. Lett. {\bf 51}, 308 (1983).

\bibitem{Jesche1}
A. Jesche, C. Krellner, M. de Souza, M. Lang, and C. Geibel; Phys. Rev. B, {\bf 81}, 134525  (2010).

\bibitem{Jesche2}
A Jesche, C Krellner, M de Souza, M Lang and C Geibel; New Journal of Physics {\bf 11}, 103050 (2009).

\bibitem{Satoh}
Kazuhiko Satoh, Toshizo Fujita, Yoshiteru Maeno, Yoshichika \={O}nukiand Takemi Komatsubara;  J. Phys. Soc. Jpn. {\bf 58}, 1012 (1989).

\bibitem{Rajan2}
V. T. Rajan ,  J. H. LowensteinandN. Andrei; Phys. Rev. Lett. {\bf 49}, 497 (1982)

\bibitem{Zlatic}
V. Zlati\'{c} and R. Monnier Phys. Rev. B {\bf 71}, 165109 (2005).

\bibitem{Miyake}
K. Miyake and H. Kohno. J. Phys. Soc. Jpn. {\bf 74}, 254 (2005).


\end{thebibliography}
\end{document}